# Synchrotron x-ray spectroscopy studies of valence and magnetic state in europium metal to extreme pressures

W. Bi[1,2,3], N. M. Souza-Neto[2,4], D. Haskel[2], G. Fabbris[1,2], E. E. Alp[2], J. Zhao[2], R. G. Hennig[5]
M. M. Abd-Elmeguid[6], Y. Meng[7], R. W. McCallum[8], K. Dennis[8], and J. S. Schilling[1]

[1]*Department of Physics, Washington University, One Brookings Dr, St. Louis, Missouri 63130, USA*
[2]*Advanced Photon Source, Argonne National Laboratory, Argonne, Illinois 60439, USA*
[3]*Department of Geology, University of Illinois at Urbana-Champaign, Urbana, Illinois 61801 USA*
[4]*Laboratorio Nacional de Luz Sincrotron, Campinas, SP 13083-970, Brazil*
[5]*Department of Materials Science and Engineering, Cornell University, Ithaca, New York 14853, USA*
[6]*II. Physikalisches Institut, Universität zu Köln, Zülpicher Str. 77, 50937 Köln, Germany*
[7]*HPCAT, Carnegie Institution of Washington, 9700 S. Cass Ave., Argonne, Illinois 60439, USA*
[8]*Materials and Engineering Physics, Ames Laboratory, Ames, Iowa 50011, USA*

**Abstract**

In order to probe the changes in the valence state and magnetic properties of Eu metal under extreme pressure, x-ray absorption near-edge spectroscopy, x-ray magnetic circular dichroism and synchrotron Mössbauer spectroscopy experiments have been carried out. The Mössbauer isomer shift exhibits an anomalous pressure dependence, passing through a maximum near 20 GPa. Density functional theory has been applied to give insight into the pressure-induced changes in both Eu's electronic structure and Mössbauer isomer shift. Contrary to previous reports, Eu is found to remain nearly divalent to the highest pressures reached (87 GPa) with magnetic order persisting to at least 50 GPa. These results should lead to a better understanding of the nature of the superconducting state found above 75 GPa and of the sequence of structural phase transitions observed to 92 GPa.

# Introduction

Whereas the majority of isolated lanthanide atoms are divalent, all but Eu and Yb are trivalent as elemental metals. That these two remain divalent is dictated by the relative stability of their half-filled ($Eu^{2+}$ - $4f^7$) or completely filled ($Yb^{2+}$ - $4f^{14}$) orbitals. The molar volumes of Eu and Yb are 40-50% larger than for the trivalent lanthanides so that both metals are highly compressible. One would thus anticipate that the application of sufficient pressure would prompt both Eu and Yb to become trivalent whereby one $4f$ electron would be promoted into the conduction band. Such a promotion would generate for Eu metal a magnetic-to-nonmagnetic transition $Eu^{2+}[4f^7 (J = 7/2)]$ → $Eu^{3+}[4f^6 (J = 0)]$ or for Yb metal a nonmagnetic-to-magnetic transition $Yb^{2+}[4f^{14} (J = 0)]$ → $Yb^{3+}[4f^{13} (J = 7/2)]$. Similar pressure-induced changes are expected to occur in many Eu and Yb compounds. In the vicinity of the so-called quantum critical point (QCP), which defines the boundary between the magnetically ordered and paramagnetic states as T→ 0 K, one would expect the emergence of unusual ground states, including exotic forms of superconductivity [1].

Unfortunately, past limitations in the available pressure range have restricted studies of such pressure-induced magnetic↔nonmagnetic transitions to relatively few systems which must be carefully doped so as to position them near a magnetic instability. An example for this is the compound EuAu$_2$Si$_2$ where the local magnetic moments on divalent Eu-ions order antiferromagnetically below 6.5 K. Under pressure the Eu ion in EuAu$_2$Si$_2$ retains its stable divalent magnetic state to at least 3 GPa [2,3]. However, the substitution of 80% of Au with Pd results in the compound Eu(Pd$_{0.8}$Au$_{0.2}$)$_2$Si$_2$ where Eu is still divalent but positioned very near a magnetic instability. Evidence for this is given by the fact that its magnetism collapses completely under only 3 GPa pressure where a mean Eu valence of 2.2 is estimated from Mössbauer spectroscopy experiments [2,3]. A significant extension of the available pressure range into the multi-Mbar region would, in many cases, eliminate the necessity to substitutionally tune the compounds under study, as for Eu(Pd$_{0.8}$Au$_{0.2}$)$_2$Si$_2$, thus avoiding the electronic structure modifications inherent to such chemical doping and enabling studies of lanthanide metals in their native form.

Modern synchrotron Mössbauer spectroscopy (SMS), x-ray absorption near-edge structure (XANES), and x-ray magnetic circular dichroism (XMCD) experiments can today be carried out in a diamond-anvil cell (DAC) to Mbar pressures, thus allowing detailed studies of magnetic↔nonmagnetic transitions and valence changes on a wide range of systems, including the elemental solids [4]. Because of their relative simplicity, elemental solids are particularly accessible to theoretical interpretation. An in-depth study of the evolution of the electronic structure and magnetic state in Eu metal under extreme pressure is thus recommended, especially in view of the recent discovery of pressure-induced superconductivity [5].

At ambient pressure Eu metal is divalent and orders antiferromagnetically at 90 K. Superconductivity in Eu for pressures above 35 GPa was predicted more than 30 years ago by Johansson and Rosengren [6] who estimated this pressure to be sufficient to push Eu to full trivalency. Recently, Debessai *et al.* [5] observed superconductivity in Eu for pressures above 75 GPa. However, the superconducting critical temperature ($T_c \approx$ 2 K) and its pressure derivative $dT_c/dP \approx$ +0.018 K/GPa are both much less than those for the trivalent *s,p,d*-electron metals Sc, Y, La and Lu which would be closely related to trivalent Eu metal [7]. These anomalously low values strongly suggest that superconducting Eu is not fully trivalent, but rather intermediate valent or divalent. This scenario is supported by the fact that to 92 GPa the structural phase transitions observed [8] in Eu under pressure (bcc→hcp→mixed-phase →primitive orthorhombic) end in a crystal structure which is *not* a member of the well-known general phase transition sequence for trivalent lanthanides under pressure hcp→Sm-type→double hcp→fcc→distorted fcc [8]. It is interesting to note that the pressure-dependent molar volume of Eu metal *V(P)* drops rapidly to 20 GPa, falling even slightly below that of the neighboring trivalent lanthanide metals Gd and Tb [9]; this result would appear to support the view that Eu metal is fully trivalent above 20 GPa pressure. However, it has been suggested that this anomalous *V(P)*-dependence may arise if Eu is in an intermediate valence state and need not imply that Eu is trivalent [10].

Theoretical estimates of the pressure necessary for a full divalent-to-trivalent valence transition in Eu range from 35 GPa [6] to 71 GPa [11]. Experimentally, the change of the valence state of Eu under pressure has been studied by several groups. Röhler [12] measured the XANES at Eu's L$_3$ edge to pressures as high as 34 GPa and reported that its valence *v* increases sharply under pressure, reaching $v \approx$ 2.5 at 10 GPa and saturating at $v \approx$ 2.64 at higher pressures. In addition, measurements of the Eu isomer shift (IS) in $^{151}$Eu Mössbauer spectroscopy studies to 14 GPa at 44 K led Farrell and Taylor [13] to conclude that Eu's valence increases rapidly under pressure, reaching $v \approx 2.45$ at 12 GPa. Interestingly, their measurement of the hyperfine field at 44 K and 12 GPa revealed that magnetic ordering was still present, in agreement with earlier resistivity studies to 15 GPa [14]. A more recent synchrotron Mössbauer experiment (SMS) at ambient temperature by Wortmann *et al.* [15] found an increase in IS to 17 GPa which was also interpreted as arising from a strong increase in Eu's valence.

It would seem clear that a reinvestigation of the evolution of the valence and magnetic states in Eu metal to higher pressures is much needed and should lead to a better understanding of the nature of the pressure-induced superconductivity. Such studies gain importance in view of speculations that the observation of a second superconducting phase in CeCu$_2$Si$_2$ under high pressure ($T_c$ = 2 K at 4 GPa) may be mediated by valence fluctuations [16].

In this paper we report results from a series of synchrotron x-ray spectroscopy experiments on Eu to pressures as high as 87 GPa, including XANES and XMCD measurements at the Eu L$_3$ edge, and SMS measurements. An analysis based on *ab initio* calculations indicates that the pressure-induced changes in XANES spectra in the 10-20 GPa range, which were previously interpreted as indicative of a marked change in Eu valence [12], originate instead from significant changes in the electronic and crystal structure at the bcc→hcp structural phase transition.

Our SMS experiments confirm that Eu's IS initially increases with pressure towards that of Eu$_2$O$_3$. However, we find that this increase most likely originates from pressure-induced changes in the properties of the *s,p,d*-electrons and is not related to a significant change in 4*f*-electron occupation, as clearly seen by the lack of a temperature-dependent Mössbauer IS between 10 K and 300 K. SMS and XMCD measurements also reveal that magnetic ordering is present to at least 50 GPa, a result that is hard to reconcile with a sizable presence of Eu$^{3+}$ ions as required for a strongly mixed-valent state. The results from the various synchrotron probes thus indicate that Eu metal remains divalent or nearly divalent to 87 GPa, a clear deviation from previous interpretations of Mössbauer and XANES data [12,13,15].

## Experimental Techniques

Due to its high reactivity, the high purity Eu sample (99.98% metal basis) from the Materials Preparation Center of the Ames Laboratory [17] was loaded into the sample chamber of the diamond-anvil cell (DAC) in an Ar glove box. The XANES experiments employed both bulk Eu samples and sandwiched Al-Eu-Al foils with a total thickness of ~ 10 μm, half being the Eu foil. The XMCD experiment was carried out on three such foils, whereas in the SMS experiment only bulk Eu was used. Unless stated otherwise, the pressure was determined *in situ* by the standard ruby fluorescence technique [18] using ruby spheres 5-10 μm in diameter [19].

High-pressure XANES experiments at Eu's L$_3$ edge were performed at beamlines 20-BM and 4-ID-D at the Advanced Photon Source (APS), Argonne National Laboratory, using transmission geometry. Undulator radiation was monochromatized with a Si(111) double-crystal and focused by a Pd toroidal mirror to ~ 120×180 μm$^2$ beam size which was further reduced to ~ 30×30 μm$^2$ by a defining slit. A Si mirror served to reject higher energy harmonic contamination in the x-ray beam. The DAC was manufactured from CuBe alloy by easyLab. To reduce the absorption from the diamond anvils, a full anvil (height ~ 2 mm) in combination with an anvil perforated down to

~100 µm wall thickness was used. Both anvils had 300 µm culet diameter beveled at 7.5 degrees to 100 µm flats and achieved pressures as high as 87 GPa in the present experiments. The Re gasket was pre-indented from an original thickness of 250 µm to 25 µm central thickness; a 50 µm diameter hole was EDM-drilled through the center to form the sample chamber.

High-pressure XMCD experiments were performed at beamline 4-ID-D using a nonmagnetic piston-cylinder mini-DAC from D'Anvils Ltd made from Inconel 718. Similar to the XANES experiment, a partially perforated anvil and a full anvil with 300 µm culet size served to reduce the absorption by the diamond anvils. However, because of the large demand for x-ray flux in XMCD experiments associated with the smaller signal-to-noise ratios than in XANES measurements, a pair of Kirkpatrick-Baez (KB) mirrors in tandem with a toroidal mirror were employed to focus the beam to ~ 12×12 µm$^2$. A superconducting magnet and He flow cryostat allowed the measurements to be carried out at 4.6 K in a 4 T magnetic field. The Re gasket was pre-indented to 50 µm with a 120 µm dia hole. The pressure was measured at ambient temperature using ruby fluorescence; the change in pressure following a cooling cycle to low temperatures was less than 3 GPa. At each pressure the DAC was first rotated at ambient temperature to minimize the Bragg peaks from the diamond anvils in the measured energy range of 6.94 - 7.0 keV. After cooling to 4.6 K, a 4 T magnetic field was applied along the photon wavevector (x-ray propagation direction) and XMCD measured in helicity switching mode. The measurement was then repeated with reversed magnetic field. At least two XMCD scans were collected for each field direction. To take into account any change in sample thickness under pressure, the XMCD amplitude is normalized by the XANES edge jump at the same pressure. XMCD data taken with opposite applied field directions were subtracted and divided by two to yield artifact-free XMCD signals.

Four SMS experiments at ambient temperature were carried out to 73 GPa pressure at beamline 3-ID-B. In the fourth experiment SMS data at low temperatures were also obtained. A standard symmetric cell (Princeton University machine shop) was used with identical full anvils with 350 µm culets beveled at 7 degrees to 180 µm flats. The Re gasket was pre-indented to 30 µm central thickness and a hole 60-90 µm in diameter EDM-drilled as a sample chamber. Ruby spheres and a small amount of Pt powder were loaded together with the Eu sample in the gasket hole. The sample pressure was determined using either the standard ruby fluorescence technique or the x-ray diffraction pattern from Pt [20] or Eu [8] as pressure marker. The x-ray beam was focused to about 10 µm in both horizontal and vertical directions using KB mirrors. In the fourth experiment, a CuBe cell from easyLab and a He flow cryostat were used to achieve low temperature and the pressure was determined *in-situ* by the ruby fluorescence technique. The x-ray beam was focused by the KB mirrors down to 10 µm in vertical and 20 µm in horizontal directions.

# Results of Experiment and Analysis

## XANES Experiments

Four different XANES experiments were carried out at ambient temperature, yielding spectra at 25 different pressures to 87 GPa. Since for pressures above ~ 30 GPa the Al-Eu-Al foil sample became too thin to permit quality data, a bulk Eu sample was used in the final experimental run to 87 GPa. Experiments to 23 GPa were carried out on a single Al-Eu-Al foil sample using as pressure medium silicone oil with a viscosity of 1000 cSt. No pressure medium was used in the other experiments to 31.3 GPa with 3 layers of foil or to 87 GPa with a pure Eu bulk sample. A comparison of the data from foil and bulk samples reveals no difference in observed changes in spectral lineshape under pressure.

A summary of the XANES results is shown in Fig. 1(a). In the bcc phase up to 12 GPa, the "white line" peak intensity (loosely defined as the main absorption feature above threshold) decreases with increasing pressure, but no emergence of a higher energy feature ~ 8.0 eV apart characteristic of the $Eu^{3+}$ state is observed, in contrast to a previous report [12] that Eu's valence increases strongly in this pressure range. As pressure is increased, it is seen that the divalent peak begins to broaden at 12.2 GPa and then separates into two distinct peaks at higher pressure. With increasing pressure, the second peak shifts to higher energy. It should also be noted that to 87 GPa the leading absorption edge at 6.97 keV does not shift significantly to higher energy. Previous XANES work also observed the emergence of two peaks for P > 10 GPa and assumed they arose from $Eu^{2+}$ and $Eu^{3+}$ components in an intermediate valence state; the mean valence was estimated from the ratio of the relative peak intensities fitted with a combination of arctan and Lorentzian functions [12]. However, this method of data analysis ignores the fact that structural phase transitions can also lead to important changes in the XANES spectra and that white line intensities can display strong pressure dependences even in the absence of valence transitions, as shown in the *ab initio* simulation in Fig. 1(b) below 12 GPa.

## Electronic Structure Calculations for XANES Spectra

To explore the effect of structural phase transitions on Eu's XANES spectra, *ab initio* simulations using both FDMNES [21] and FEFF8 [22] were carried out. The self-consistent calculations treated the seven 4*f* electrons of $Eu^{2+}$ as core electrons, thus holding their occupancy fixed, and used the known pressure-induced structural sequence [bcc→hcp at 12 GPa, mixed phase 18-65 GPa, orthorhombic for 65-92 GPa] [8]. Fig. 1(b) shows the results of the FEFF8 simulation to 75 GPa. The simulated spectra shown in the figure were artificially shifted in energy to match at their thresholds. The calculated changes in XANES spectra with pressure are seen to bear a clear resemblance to the measured spectra in Fig. 1(a). In the bcc phase the calculated spectral lineshape changes only little with pressure, but the white line peak intensity is strongly suppressed. At the bcc-hcp phase transition at 12 GPa the white line peak becomes broadened and a second peak appears and shifts to higher energy with increasing pressure.

To check the assumption in the FEFF8 *ab initio* calculations that the number of 4*f* electrons remains fixed at seven, density functional theory (DFT) calculations to 87 GPa using the LDA+U approach setting U = 7 eV were carried out, and their results used as input for XANES simulations using the FDMNES code. The all-electron DFT calculations are performed using the linearized augmented plane wave method and the PBE exchange-correlation functional [23] implemented in the WIEN2K code (Version 11) [24]. The parameters of the LAPW basis and the Brillouin-zone integration are chosen to achieve a total-energy accuracy of 1 meV/atom. This requires a muffin-tin radius $R_{MT}$ = 2.3 $a_o$ and a value of $R_{MT} \cdot k_{max}$ = 8.0, where $k_{max}$ is the plane wave cutoff for the basis functions. The Brillouin-zone integration is performed using a mesh corresponding to about 2000 k-points in the full Brillouin zone. Since Eu orders in a non-collinear, antiferromagnetic state at ambient pressure, resulting in a low symmetry magnetic structure difficult to treat in DFT, a ferromagnetic ground state was assumed instead. The U parameter used in these simulations places 4*f* electrons at about 1 eV below the Fermi level, in agreement with the calculations of Turek *et al.* [25]. This placement is more likely to induce a 4*f* valence transition than the 3 eV reported by Kuneš *et al.* [26].

Fig. 2(a) shows the predicted changes in *f*, *d*, *p*, *s*, and interstitial electron occupancy $\Delta n$ from DFT using the x-ray derived structural data for Eu metal to 92 GPa from Ref. 8. The *d*-electron occupancy increases rapidly with pressure. In contrast, the *f*-electron occupancy decreases only by about 0.14 electrons from ambient pressure to 20 GPa, increasing again at higher pressure to nearly its initial value; the changes in *p* and *s* occupancies are even smaller. The increase in the number of *d* electrons is seen to arise principally from the interstitial electrons. This increase in the *d* electron occupancy implies that the number of empty *d* states must decrease which explains the strong decrease in the white line peak intensity under pressure observed in Fig. 1, especially in the bcc phase. The increase in *d* electron occupancy with pressure is due to the well known *s-d* transfer, a hallmark of the change in electronic properties with pressure across much of the periodic table where the number of *d* electrons per atom plays a central role in determining both the binding energy and the most favorable crystal structure.

In Fig. 2(b,c) the predicted changes in *f*, *d*, *p*, *s*, and interstitial electron occupancy $\Delta n$ is shown assuming a single structure (bcc or orthorhombic *Pnma*) for all pressures. It is seen that the predicted changes in $\Delta n$ do not depend sensitively on the crystal structure assumed. This fact will be useful in a later section where the pressure-dependent contact electron density and Mössbauer isomer shift are calculated.

In Fig. 3(a) it is seen that once Eu enters the hcp phase at 12 GPa the DFT-based, FDMNES simulated spectra show a double peak feature which is consistent with both the experimental data and the *ab initio* FEFF8 simulation in Fig. 1(b). In view of the small change in *f* occupancy to 20 GPa and the large change in density of states seen in the DFT calculations at the bcc→hcp transition, together with the two *ab initio* simulations presented for the different structural ground states, we suggest that the most likely scenario for the changes in XANES spectra observed in the present high-pressure experiment arise predominately from changes in crystal structure (and associated changes in electronic structure) without invoking any significant change in Eu valence.

In Fig. 3(b) we show the total magnetic moment per Eu ion calculated in the LDA+U scheme. This moment is seen to increase by about 0.1 $\mu_B$ at the bcc-hcp transition, primarily arising from the 5*d* electrons. However, the overall tendency is that the total moment decreases slowly under pressure at the approximate rate -0.004 $\mu_B$/GPa to a value of approximately 6.9 $\mu_B$ at 100 GPa, still a sizeable value. This simulation thus indicates that magnetic order in Eu survives into the Mbar pressure range with the caveat that a ferromagnetic state was used in the DFT calculations at all pressures.

We thus find no clear signature of $Eu^{3+}$ in the present XANES data within the bcc phase (to 12 GPa), in disagreement with previous work [12]. If one were to be unaware of the results from the present XANES experiments and simulations presented above, one might be led to conclude that the "doublet" peaks appearing above 12 GPa in Fig. 1(a) give evidence for coexisting $Eu^{2+}$ and $Eu^{3+}$ states. Above 12 GPa the line shapes in the present spectra are similar to those reported earlier [12]. However, the present data fail to confirm the presence of a second peak in the bcc phase. We also note that the two peaks at 17 GPa in Fig. 1(a) are not separated by 8 eV, as is well established for $Eu^{2+}$ and $Eu^{3+}$ states, but rather only by 6.5 eV. This value is obtained from fitting the data with two arctan and two Lorentzian functions. With increasing pressure the peak separation increases to almost 9 eV at 87 GPa. Also, the appearance of the peak "doublet" occurs concomitant with the bcc→hcp transition. If this were related to a valence change, one would have to conclude that the valence transition is rather sharp and confined to the 10-20 GPa pressure range, becoming complete at 23 GPa where the XANES spectra (and the relative intensities of the doublet peaks) cease to show significant change to higher pressures. This is fundamentally different from the conclusions of previous XANES work [12] where it is reported that the valence increases rapidly up to 2.5+ within the bcc phase, saturating at 2.64+ for pressures to 34 GPa.

Because of the complexity of the changes in Eu's XANES spectrum with pressure due to the series of structural phase transition occurring, it is desirable to test the above conclusions using alternative diagnostic tools such as XMCD, which probes the strength of the sample's magnetization, and SMS which gives information on changes in both the valence and the magnetization on a local scale.

## XMCD Experiments

To further explore the changes occurring in the valence and magnetic states of Eu metal under pressure, XMCD experiments were carried out to pressures as high as 59.4 GPa at 4.6 K. Since XMCD only occurs in the presence of non-zero element-specific magnetization, and since Eu is antiferromagnetic at ambient pressure, a net magnetic moment must be induced by applying a strong magnetic field to cant the Eu spins. Due to the lack of single-ion anisotropy associated with the half-filled $4f^7$ configuration of $Eu^{2+}$, this is easily achieved with a 4 T magnetic field [27].

XANES and XMCD spectra were measured at 4.6 K for pressures 3.3, 7.1, 10.3, 15, 21.4, 30.7, 40, 49.1 and 59.4 GPa. As seen in Fig. 4(a), the low temperature XANES data for 3.3 and 21.4 GPa agree well with those in Fig. 1(a) from the previous experiments at ambient temperature. The fact that the XANES spectra are independent of temperature suggests that Eu is *not* in an intermediate valence state [28]. The XMCD signal amplitudes at 3.3 and 21.4 GPa are shown in Fig. 4(b); note the strong increase in signal amplitude at 21.4 GPa.

The dependence of the XMCD amplitude on pressure is shown in Fig. 5 to 59.4 GPa. The error bars for the pressure give the change in pressure at ambient temperature before and after cooling, the latter being higher. The uncertainty in the XMCD signal represents the standard deviation of the XMCD amplitude from different scans. The canted magnetism is seen to stay nearly constant in the bcc phase below 12 GPa. At higher pressures the magnetism first increases sharply up to 21.5 GPa before decreasing gradually at higher pressures. The shape of the dependence in Fig. 5 bears some resemblance to the simulated change in total magnetic moment under pressure from Fig. 3(b); this suggests that the change in Eu's magnetism may arise at least in part from the pressure-induced structural phase transitions.

As seen in Fig. 5, the normalized XMCD amplitude in the bcc phase ~ 0.5(3)% is non-zero, a result of canting of antiferromagnetically ordered Eu moments in the 4 T applied field. Magnetization data on Eu single crystals show a 4 T field-induced ~ 1 $\mu_B$/Eu atom [29]. The XMCD amplitude reaches a value of about 2.6(5)% near 21 GPa, a five-fold increase relative to ambient pressure. This value is comparable to the ~ 3% XMCD amplitude observed in a fully saturated EuO sample at 9 K and ambient pressure where Eu ions are clearly divalent [30]. We note that the 2.6(5)% XMCD amplitude at 21 GPa is 20 times larger than the 0.13% amplitude measured in the Van Vleck paramagnet $Eu_2O_3$ at 4 T and 9 K at ambient pressure where Eu is trivalent [29]. The strength of XMCD data in Fig. 5 is consistent with divalent Eu and clearly indicates that Eu does not become trivalent in the measured pressure range.

To gain information on the nature of the magnetic ordering in Eu under pressure, the field dependence of the raw (unnormalized) XMCD amplitude was measured at 4.6 K at the resonant energy of 6.972 keV which corresponds to the maximum XMCD amplitude. At ambient pressure the dependence of the XMCD amplitude is linear (see Fig. 6), as expected for field-induced canting of Eu's antiferromagnetic state. However, as seen in Fig. 6, at 21.4 GPa the signal begins to saturate for magnetic fields above 2 T. The same holds true for measurements at 30.7 and 49.1 GPa. This saturation gives evidence that at pressures of 21.4 GPa and above the nature of the magnetic order in Eu has changed. A possible explanation is a pressure-induced transition from an AFM to a FM structure, which would be consistent with the sharp rise in the XMCD amplitude above 15 GPa seen in Fig. 5. The lack of significant hysteresis and remanent magnetization in the high pressure data, however, may indicate that this description is too simplistic. DFT calculations of the magnetism of Eu metal at high pressures should help clarify the nature of the observed changes with pressure in the response of Eu moments to an applied field.

Here we would like to mention that the occurrence of a pressure-induced magnetic phase transition at 21.4 GPa and above requires the existence of relatively stable, localized $Eu^{2+}$ magnetic moments, which is fully consistent with the XANES results (see above). In fact, the enhancement of exchange interactions above 15 GPa is inconsistent with a significant presence of $Eu^{3+}$ ($J = 0$) states. An $Eu^{2.6+}$ intermediate valence state, as postulated in previous XANES work [12], would be expected to strongly suppress exchange interactions, contrary to observation.

## Synchrotron Mössbauer Spectroscopy Experiments

### Isomer Shift (IS) at Ambient Temperature

SMS data on $^{151}$Eu metal were collected in three separate high-pressure experiments at ambient temperature: run 1 (4, 9, 14, 17, and 25 GPa), run 2 (12, 19.9, 24, 29, and 35.6 GPa), and run 3 (23.1, 45.2, 52.4, 58.0, 65.5, and 72.6 GPa), all shown in Fig. 7 (a,b,c). In each of the measurements the Mössbauer spectrum from the Eu sample alone was first measured to check the sample purity. The pure exponential decay of the signal indicates the absence of $Eu^{3+}$ impurities. Both reference and Eu samples were measured to obtain the isomer shift (IS). The reference was then removed from the x-ray beam path and x-ray diffraction carried out on the Eu sample without moving its position. In run 1 at low pressures, $Eu_2O_3$ with IS =1.024 mm/sec served as the reference. However, under pressure the IS of the $^{151}$Eu sample increases from its ambient pressure value of -7.33 mm/sec [13] and approaches that of the reference, thus yielding fewer quantum beats in the SMS spectra and making an accurate determination of the IS difficult. Therefore, at pressures of 25 GPa and above, EuS was used as reference where IS = -11.496 mm/sec. The Mössbauer IS was extracted by fitting the data with CONUSS software [31,32]. All $^{151}$Eu SMS spectra obtained are shown in Fig. 7 (a,b,c) in both time and the corresponding energy domains.

The pressure-dependent IS for Eu at ambient temperature from all three experimental runs is plotted in Fig. 8 and compared to the results of previous studies by Farrell and Taylor [13] at 44 K to 12 GPa and Wortmann *et al.* [15] at ambient temperature to 22 GPa. The error bars in pressure are given by the change in pressure measured before and after each SMS measurement and the error bars in the isomer shift reflect the refinement uncertainty in the data fits. The present IS data agree well with previous studies [13,15], particularly for the data of Farrell and Taylor to 12 GPa. The application of 20 GPa pressure is seen to cause IS to increase rapidly by almost 6 mm/sec. However, between 25 and 30 GPa IS decreases abruptly followed by a slow decrease to the maximum pressure of the experiment 73 GPa. We note that the increase and/or decrease of the IS reflects corresponding changes in the *s*-electron density at the Eu nucleus.

From Fig. 8 the initial rate of increase of IS with pressure to 10 GPa at either 44 K or ambient temperature is given by $d\mathrm{IS}/dP = +0.35$ (mm/s)GPa$^{-1}$, in approximate agreement with earlier experiments by Klein *et al.* [33] to 1.6 GPa who found $d\mathrm{IS}/dP = +0.48(5)$ (mm/s)GPa$^{-1}$. Both Farrell and Taylor [13] and Wortmann *et al.* [15] have interpreted this increase as due *solely* to a change in Eu's valence $\Delta v$ with pressure and estimated $\Delta v$ by extrapolating linearly between the difference $\Delta\mathrm{IS} \approx 10$ mm/s in IS values at *ambient pressure* between trivalent Eu$^{3+}$ and divalent Eu$^{2+}$ in typical metallic systems. Since to 14 GPa $\Delta\mathrm{IS} = 4$ mm/s, the valence change estimated in this way would then be $\Delta v = 4/10 = 0.4$ which would place Eu at 14 GPa squarely in the intermediate valence range with $v = 2.4+$.

Such a simple analysis relates the pressure-induced change in IS *solely* to a corresponding change of the number of 4*f*-electrons. In reality, however, the increase of IS (or the *s*-electron density at the nucleus) with pressure cannot be fully attributed to a true valence shift towards Eu$^{3+}$ unless the volume effect due to isothermal compression is corrected. Moreover, one has to consider relevant mechanisms other than valence change which also can lead to a comparable increase in IS [33, 34]: (i) compression of *s*-like conduction electrons, mainly 6*s* electrons, or (ii) increase of the intra- and interatomic exchange interactions involving the 4*f*, 5*d*, 6*p*, and 6*s* electrons. The importance of the contributions (i) and (ii) to the increase of IS with pressure have been estimated for divalent Eu in EuO [33] and EuAl$_2$ [34]. For example, the valence of Eu in EuAl$_2$ remains in a stable Eu$^{2+}$ state to 41 GPa, even though the change in IS to 41 GPa (30% volume decrease) is about 7 mm/s [35].

It is enlightening to compare the initial change in IS under pressure for Eu to that for Eu-compounds where Eu is known to be in a stable divalent state. Because Eu is approximately *ten times* more compressible ($B_o = 11$ GPa [8]) than typical Eu-compounds ($B_o \approx 110$ GPa) [36], it is advisable to compare the changes in IS not as a function of pressure but rather as a function of change in relative sample volume $\Delta V/V_o$. Since from Fig. 8 the increase in the IS for Eu metal is $\Delta\mathrm{IS} = +5.6$ mm/s at 20 GPa pressure, where the relative volume decreases by $\Delta V/V_o = -0.55$ [8], it follows that $\Delta\mathrm{IS}/(\Delta V/V_o) = -10$ mm/s. This value of $\Delta\mathrm{IS}/(\Delta V/V_o)$ has the same sign, but is much lower than typical values for Eu intermetallic compounds where Eu is in a stable divalent state throughout [3,37]! In EuAu$_2$Si$_2$, for example, it is found that $\Delta\mathrm{IS}/(\Delta V/V_o) = -31$ mm/s [3] and the IS is temperature independent. We note that in the related compound Eu(Pd$_{0.8}$Au$_{0.2}$)$_2$Si$_2$, the IS is strongly temperature dependent, the hallmark of an intermediate valence system; here the change in IS under pressure is found to be $\Delta\mathrm{IS}/(\Delta V/V_o) = -115$ mm/s [3], a value 15× larger than that found for pure Eu metal! It is thus highly unlikely that the increase in IS for Eu to 20 GPa seen in Fig. 8 arises from a change in the valence of Eu. Furthermore, the decrease in IS for pressures above 20 GPa seen in Fig. 8 is almost certainly not related to a change in valence. If Eu's valence were to change under pressure, it would be expected to monotonically increase, except perhaps at those pressures where structural phase transitions occur.

From the above it appears all but certain that the increase in Eu's IS to 20 GPa seen in Fig. 8, which arises mainly from an increase in the *s* electron density at the Eu nucleus $\rho_0$, is not due to a pressure-induced increase in Eu's valence. At ambient pressure the contribution of the 6*s* conduction electrons to the IS of the $Eu^{2+}$ configuration is estimated to be S ≈ 6 mm/s [34]; at 20 GPa the molar volume of Eu metal decreases by roughly a factor of two [8]. Assuming that under pressure the 6*s* electron density increases uniformly with decreasing volume as $\rho_0 \propto V^{-1}$, the 6*s* electron contribution to the enhancement of IS under 20 GPa pressure can be estimated to be: ΔS ≈ (+6 mm/s)(V(0)/V(P)) ≈ (+6 mm/s)(2) = +12 mm/s, a value more than double that (+5.3 mm/s) observed in experiment (see Fig. 8). The Mössbauer IS data thus clearly support the view that Eu metal remains nearly divalent to 20 GPa. The anomalous decrease in IS to pressures from 20 to 73 GPa is likely related to the structural transitions that occur, including mixed phase regions. But clearly, as emphasized above, the decrease in IS above 20 GPa is not consistent with an increase in Eu's valence since such an increase should lead to a strong further enhancement in the IS.

In this context, it is interesting to compare the observed anomalous pressure dependence of IS with our pressure-dependent DFT, regarding the variation in occupation numbers of the different electronic states with pressure shown in Fig. 2. The results indicate that the *d*-electron occupation number increases rapidly with pressure, in contrast to the very small change in the *s*, *p* and *f* occupations. There are two arguments that the observed pressure dependence in IS cannot originate from an increase of the *d* occupation: (a) an increase of the occupation number of the *d* electrons with pressure alone would lead to a significant decrease of $\rho_0$ (or IS) due to increased screening of the *s*-electrons, which is the opposite of our experimental observation; (b) the monotonic increase of the *d* occupation does not follow the observation that the IS first increases up to about 20 GPa and then slightly decreases. The results of DFT calculations on Eu are presented below and shed light on the origin of the pressure-dependent IS in Fig. 8.

## Isomer Shift and Hyperfine Field at Low Temperature

Unequivocal information on possible changes in the valence of Eu metal under pressure is given by determining the degree to which the IS changes with decreasing temperature [37, 38, 39]. Below the magnetic ordering temperature, the SMS spectrum is split by the hyperfine field. SMS measurements on Eu to extreme pressures and low temperatures give definitive information on what changes occur in both Eu's valence and magnetic state as a function of pressure.

To investigate the temperature dependence of IS, a separate experiment was carried out in a membrane-driven DAC made of CuBe alloy (easyLab). As seen in Fig. 8, in run 4 the IS at ambient temperature was determined for 3.5, 8.5, 21.9 GPa and at low temperatures for 8.3 GPa at 11 K and for 18.6 GPa at 115 K. It is seen that in this pressure range the IS is independent of temperature! This is clear evidence that to 22 GPa Eu remains nearly divalent and has not entered the intermediate valence state. The temperature independence of the IS for Eu to 14 GPa is also evident in Fig. 8 from a comparison of our data from runs 1 and 4 to the earlier data of Farrell and Taylor [13]. In contrast, for a true intermediate valence compound, like $EuPd_2Si_2$, the IS decreases as the temperature is lowered from 300 K to 4 K by approximately 6 mm/s [40].

The SMS spectra of Eu at low temperature were also measured during warming at selected pressures, without using a reference sample. As shown in Fig. 9, for 20.2 GPa at 11 K as well as for 19.7 GPa at 81 K, Eu still orders magnetically, as inferred from the oscillations in the spectra. On the other hand, on warming up from 81 K to 115 K at 18.6 GPa, the oscillations disappear, proving that Eu no longer orders magnetically at 115 K. At the highest pressure measured in run 4 (27.7 GPa), Eu still orders magnetically at a temperature well above 11 K, in agreement with the earlier resistivity measurements by Bundy and Dunn [14] and the present XMCD results shown in Fig. 5.

An attempt was made to estimate the magnetic hyperfine field by fitting the split SMS data at temperatures below the ordering temperature. However, it was not possible to obtain an acceptable fit either with or without quadrupole splitting, possibly because Eu may be in a mixed crystalline phase at these pressures, as is the case at ambient temperature [8]. In order to obtain a satisfactory fit to the data, the first step is to establish the crystal structure at high pressures and low temperatures through x-ray diffraction experiments. To learn how the magnetic hyperfine field changes under pressure and whether magnetic ordering persists in the pressure range above 75 GPa where superconductivity appears, further low temperature experiments on Eu metal are clearly needed.

## DFT Calculations of the Isomer Shift

The IS is proportional to the electron density at the nucleus, the so-called contact electron density $\rho_0$. Only the *s*-electrons contribute to $\rho_0$. To accurately determine the contact density, we perform all-electron DFT calculations using the linearized augmented plane wave (LAPW) method and the PBE exchange-correlation functional [23] implemented in the WIEN2K code (Version 11) [24], as discussed in an earlier section of this paper. For the high-pressure orthorhombic *Pnma* phase, the lattice parameters were determined using the VASP program (Vienna *ab initio* simulation program) employing the projector augmented wave method within the frozen-core approximation [41, 42]. The VASP calculations are performed describing the [Kr]$4d^{10}$ electrons as frozen core electrons and a plane wave cutoff energy of 600 eV and a *k*-point density of 50 per Å$^{-1}$ ensure an accuracy of 1 meV/atom.

Figure 10 shows the change in contact density $\Delta\rho_0$ as a function of (a) volume and (b) pressure from the all-electrons DFT calculation for both ferromagnetic and antiferromagnetic Eu in the bcc and *Pnma* structures. The change contact density $\Delta\rho_0$ is given relative to the value of the ferromagnetic bcc phase at zero pressure. Since a detailed description of the nuclear γ transition of Eu is still outstanding, we here assume the usual linear relationship between the isomer shift IS and the change in contact density $\Delta\rho_0$: IS = A·$\Delta\rho_0$ + C [43]. The right ordinate in Fig. 10 illustrates the IS using the values A = 0.21 and C = −7.3 where the units of IS and $\Delta\rho_0$ are, respectively, mm/s and electrons/(Bohr radius)$^3$.

Up to about 10 GPa the contact density $\rho_0$ is primarily a function of atomic volume. At higher pressure, the contact density deviates between the various structures and magnetic orderings. However, in all cases $\rho_0$ rapidly increases as a function of pressure and reaches a maximum around about 30 GPa and then slowly decreases. This behavior closely reproduces the salient features of the measured pressure-dependent IS data in Fig. 8; a rapid rise followed by a gradual fall as pressure is increased.

The changes in the contact density are dominated by the changes in the occupancy of the *s* valence electrons. Fig. 2 (b and c) shows the changes in electronic occupation of states with *s*, *p*, *d*, and *f* character surrounding the Eu atoms in the ferromagnetic bcc phase and the antiferromagnetic *Pnma* phase, respectively. The 6*s* state of the Eu atom is nearly fully occupied and the changes in occupation of the *s*-electron states are quite small. Figure 11 compares the change in *s*-electron occupation $\Delta n_s$ with the change in contact density $\Delta\rho_0$ due to the valence electrons for the ferromagnetic bcc phase. A significant fraction of the change in contact density $\Delta\rho_0$ is caused by the small change in *s* occupation $\Delta n_s$. The remaining change in contact density $\Delta\rho_0$ may be due to the compression of the *s*-orbitals. The DFT calculations for the high-pressure *Pnma* phase show that the antiferromagnetically ordered state has a lower enthalpy than the ferromagnetic state. The electronic structure of both the ferromagnetic bcc and the antiferromagnetic *Pnma* phases are dominated by an *f* to *d*-electron transfer (see Fig. 2). In addition to the observed *f* to *d*-electron transfer under pressure, the antiferromagnetically ordered *Pnma* phase also displays a partial reduction in local moment. The local *f*-electron spin moment is reduced under pressure from 6.6 to 6.1 $\mu_B$. About a third of this reduction is caused by an increase in the occupation of the minority *f*-electron states. This observed partial reduction of local moment may indicate the importance of spin-fluctuations in this system.

## Discussion

In contrast to previous reports that the valence of Eu in Eu metal increases sharply under pressure to 2.6+ at 20 GPa [12,13,15], the present high-pressure XANES, XMCD and SMS experiments concur that Eu's valence state remains close to divalent at 20 GPa. The present XANES measurements, in fact, indicate that Eu remains nearly divalent to 87 GPa. The splitting of the XANES peak above 12 GPa is shown by both *ab initio* and DFT simulations to result naturally from the bcc→hcp structural transition for divalent Eu. The previous XANES studies [12] failed to take the structural transition into account and identified the peak splitting as arising from a sizable $Eu^{3+}$ component and thus a significant increase in valence.

Whereas the present SMS studies show that the increase in the Mössbauer isomer shift $\Delta IS$ to 20 GPa can be fully accounted for by an increase in the *s* contact electron density $\rho_0$, this simple effect was neglected in the previous studies [13,15] which ascribed the measured $\Delta IS$ to a change in 4*f* electron count alone (change in valence). As pointed out above, however, the measured dependence of the IS of Eu metal on sample volume is far less than that for well known intermediate valence systems. In addition, the fact that Eu's IS is independent of temperature at 8.3 and 18.6 GPa can only be understood if Eu's valence remains near 2+ at these pressures. Since above 20 GPa the IS of Eu *decreases* monotonically to 73 GPa, the opposite direction anticipated were Eu's valence to increase, one must conclude that the present SMS studies support that Eu metal remains essentially divalent to 73 GPa and not intermediate valent.

The present SMS and XMCD studies find magnetic ordering in Eu metal to at least 27.7 GPa and 50 GPa, respectively, at temperatures well above 10 K. In fact, recent electrical resistivity studies by Tatsukawa and Shimizu [44] indicate that magnetic order in Eu may persist to pressures as high as 90 GPa. This is only possible if Eu remains nearly divalent to these pressures. A similar result was found in previous Mössbauer effect studies across the $Eu(Au_{1-x}Pd_x)_2Si_2$ series [3]. In these studies the compound $EuAu_2Si_2$ was found to order antiferromagnetically below 6.5 K and exhibit Curie-Weiss behavior at higher temperatures with an effective moment very near that calculated for the free $Eu^{2+}$ ion [3]. On the other hand, $EuPd_2Si_2$ is a well known intermediate valence compound which does not order magnetically. As the Pd concentration *x* in $Eu(Au_{1-x}Pd_x)_2Si_2$ is increased from *x*=0 to *x*=1, the compound moves from a stable divalent to an intermediate valence state. For $x = 0.8$, the Eu ion in the compound $Eu(Au_{0.2}Pd_{0.8})_2Si_2$ is still essentially divalent, as evidenced by its full free $Eu^{2+}$-ion Curie-Weiss temperature and antiferromagnetic ordering at 32 K, but lies near a valence instability. This is evidenced by the fact that its Néel temperature $T_N$, hyperfine field $B_{eff}$, and isomer shift IS change sharply with pressure, whereas for $EuAu_2Si_2$ these quantities show little pressure dependence to 3 GPa. A detailed analysis reveals that whereas $EuAu_2Si_2$ retains a stable divalent magnetic state to 3 GPa, Eu's valence in $Eu(Pd_{0.8}Au_{0.2})_2Si_2$ increases under pressure, reaching $v \approx 2.2+$ at 3 GPa. Most significant for the present experiments on Eu metal is the fact that already at 1.2 GPa the Néel temperature for $Eu(Pd_{0.8}Au_{0.2})_2Si_2$ has decreased from 32 K at ambient pressure to well below 4 K so that magnetic order and intermediate valency are found to only coexist for values of the valence $v < 2.2+$ [3]. The fact that in the present XMCD experiments the magnetic order in Eu metal survives at temperatures well above 10 K to pressures as high as 50 GPa supports our conclusion that to this pressure the Eu ion remains essentially divalent.

The equation of state of Eu metal has been recently determined to 92 GPa [8]. The molar volume of divalent Eu is found to decrease from ~29 cm$^3$/mol at ambient pressure to approximately half this value at 20 GPa where it falls slightly below the published molar volume for the neighboring trivalent rare-earth metal Gd [45]. This would appear to support the view that at 20 GPa Eu metal is essentially trivalent. However, in Fig. 2 it can be seen that for Eu metal the increase in the *d*-electron count per ion $N_d$ is large; in fact, to 20 GPa the increase in $N_d$ is nearly twice that for Gd metal, its trivalent neighbor in the rare-earth series [46]. This would be expected to lead to an anomalously strong decrease in the molar volume of Eu metal since, as is well known, *d* electrons make the dominant contribution to the binding energy and determine the type of crystal structure in transition-metal-like (*d*-electron) systems, like the rare-earths. In fact, many years ago Duthie and Pettifor [46] demonstrated that the value of the single parameter $N_d$ is able to account for the observed changes in crystal structure across the rare-earth series both at ambient and high pressures. We also note that important contributions to the binding energy may arise if Eu metal is in or near an intermediate valence state at this pressure [10]. The low molar volume for $P \geq 20$ GPa is thus not necessarily inconsistent with Eu metal remaining in a divalent state over the pressure range of the present experiments.

Future SMS experiments are planned to low temperatures under pressures near 100 GPa in an effort to establish the valence and magnetic states of Eu metal in the region of pressure and temperature where superconductivity appears [5].

**Acknowledgements.** Work at both Washington University and the Advanced Photon Source (APS) was supported by the Carnegie/DOE Alliance Center (CDAC) through NNSA/DOE grant number DE-FC52-08NA28554, the Center for Materials Innovation, and the National Science Foundation through grant number DMR-0703896. Work at Argonne is supported by the U.S. Department of Energy (DOE), Office of Science, Office of Basic Energy Sciences, under Contract No. DE-AC-02-06CH11357. Work at Cornell was supported by by the National Science Foundation under Award Number CAREER DMR-1056587, and by the Energy Materials Center at Cornell (EMC2), funded by the U.S. Department of Energy, Office of Science, Office of Basic Energy Sciences under Award Number DE-SC0001086. This research used computational resources of the Texas Advanced Computing Center under Contract Number TG-DMR050028N and of the Computation Center for Nanotechnology Innovation at Rensselaer Polytechnic

Institute. The authors would like to thank Abdel Alsmadi for experimental support, Wenge Yang for providing Be gaskets as well as Mali Balasubramanian and Steve Heald for user support in the XANES experiment at 20-BM. Helpful communications with G. Wortmann and J. Röhler are also acknowledged.

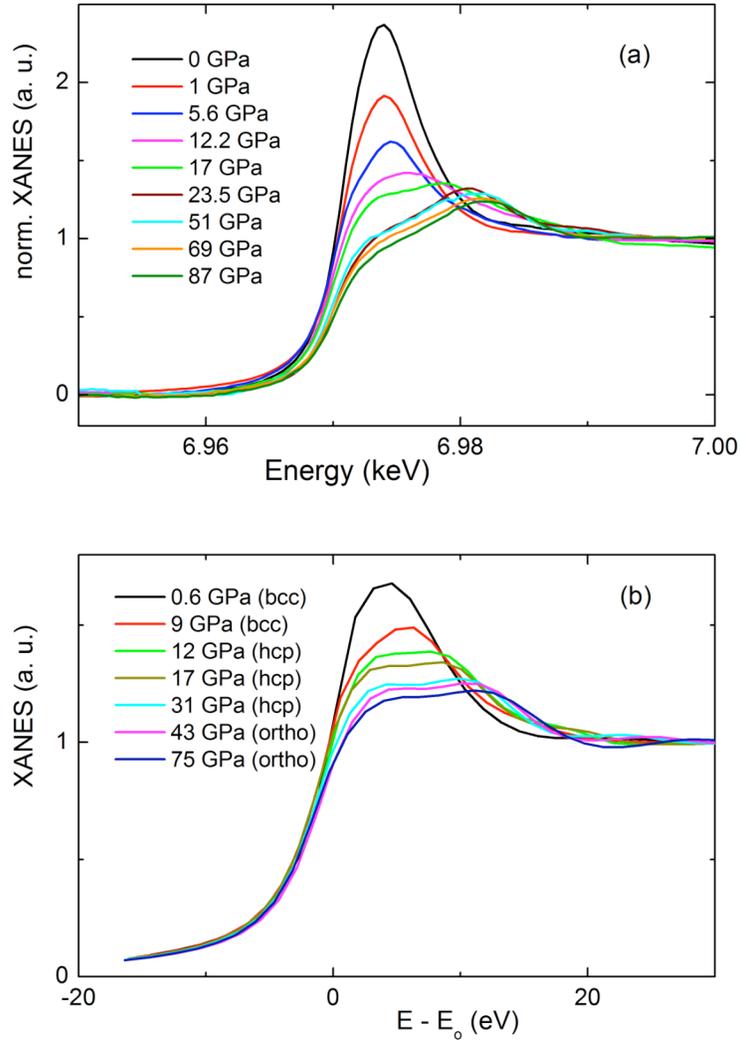

Fig. 1. (Color online) (a) XANES spectra at ambient temperature for Eu at selected pressure to 87 GPa; (b) *ab initio* simulation of XANES spectra FEFF8 under pressure to 75 GPa (see text). $E_o$ is the absorption threshold for the $L_3$ absorption edge.

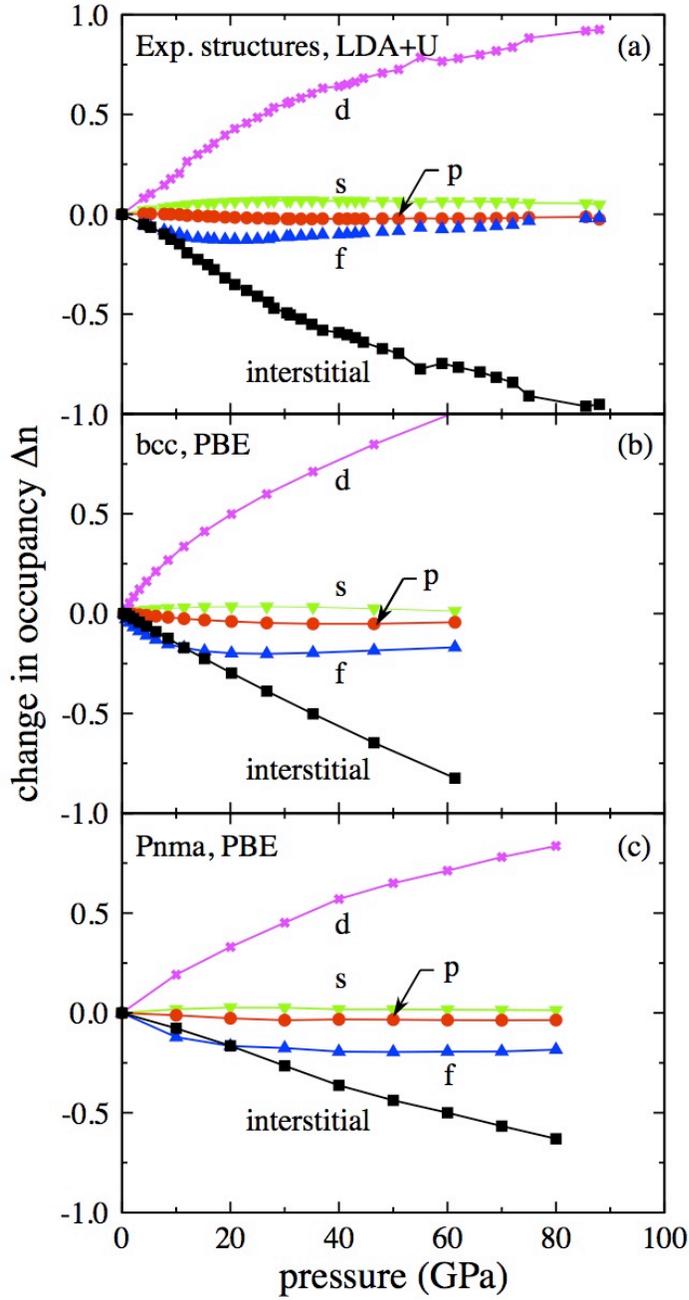

Fig. 2. (Color online) (a) DFT simulations in LDA+U scheme for Eu showing change in electron occupancy $\Delta n$ under pressure for $s, p, d, f$ and interstitial electrons. The simulations were based on the crystal structure data from Ref. 8: bcc from 0 to 11 GPa, hcp from 11 to 35 GPa, *Pnma* (orthorhombic) from 36 to 88 GPa. (middle and lower panels) PBE simulations for ferromagnetic bcc and antiferromagnetic orthorhombic *Pnma* structures, respectively. For (a,b,c) panels the electron occupancy $n$ at P = 0 equals (2.07, 2.00, 1.99) for $s$, (5.54, 5.27, 5.27) for $p$, (0.12, 0.21, 0.21) for $d$, (6.74, 6.60, 6.60) for $f$ and (2.53, 2.93, 2.92) for interstitial electrons, respectively.

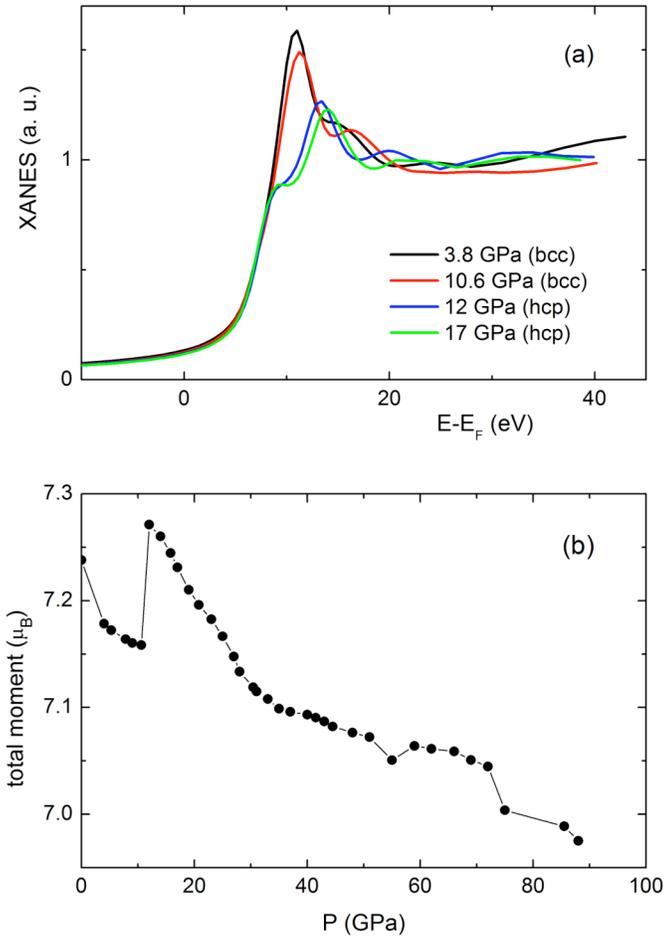

Fig. 3. (Color online) DFT simulations in the LDA+U scheme for Eu show: (a) XANES spectra FDMNES in bcc and hcp phases, $E_F$ is the Fermi energy, (b) total magnetic moment per Eu ion in units of Bohr magneton $\mu_B$.

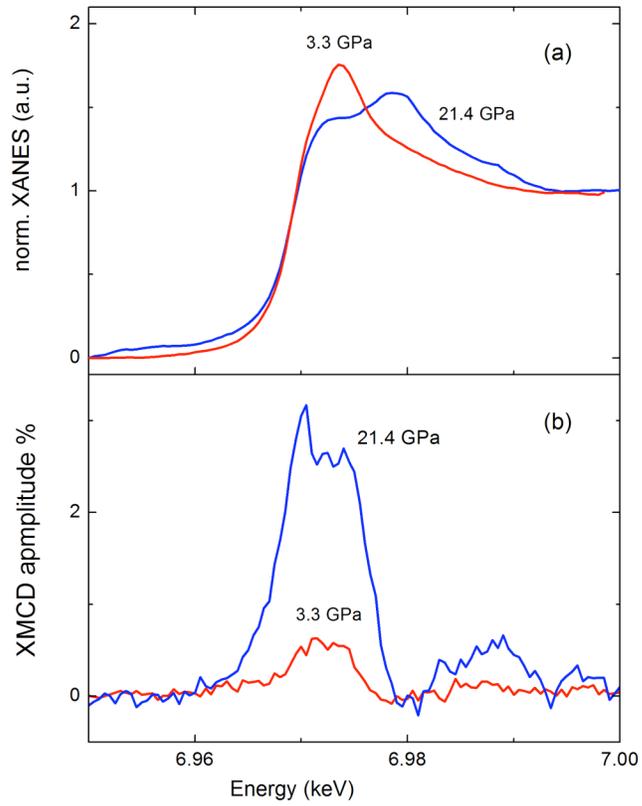

Fig. 4. (Color online) (a) Normalized XANES spectra, and (b) corresponding XMCD amplitudes at 3.3 and 21.4 GPa.

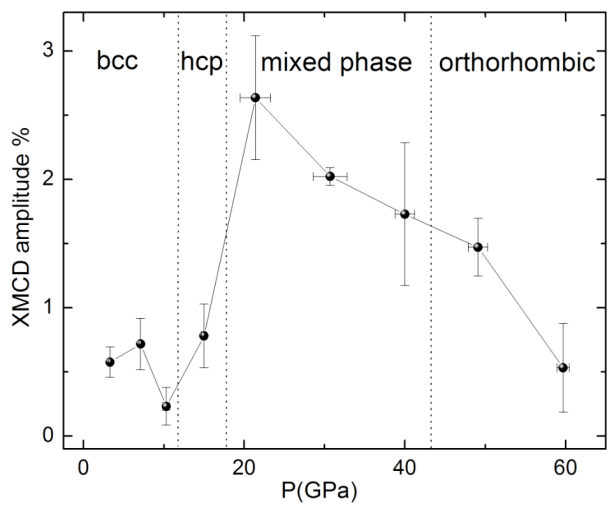

Fig. 5. XMCD amplitude of Eu metal under pressure to 60 GPa. Structural phase boundaries from Ref. 8.

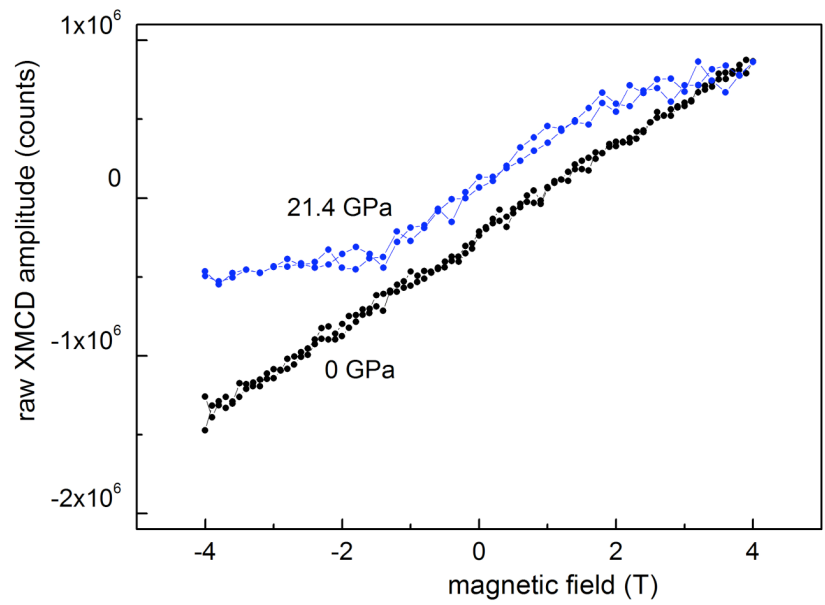

Fig. 6. (Color online) Magnetic field dependence of the XMCD amplitude at 4.6 K for a Al-Eu-Al foil sample outside the DAC at ambient pressure, and at 21.4 GPa in the DAC.

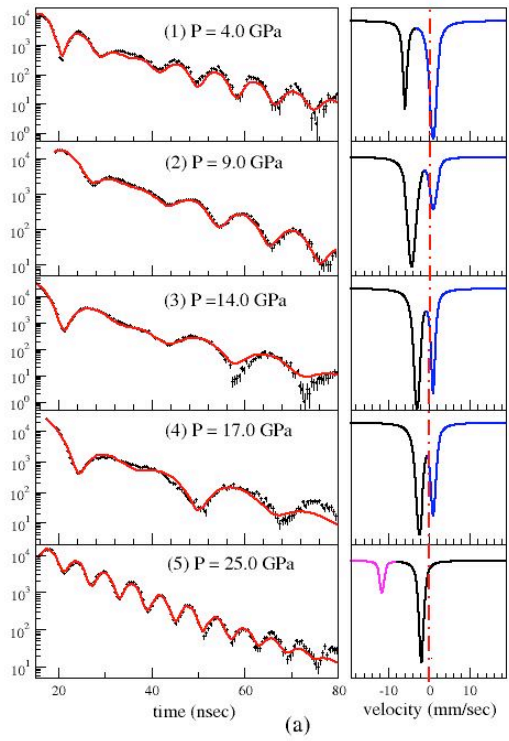

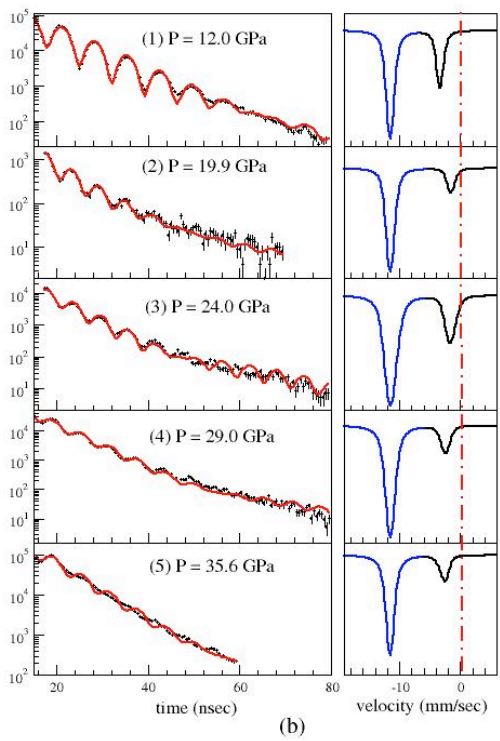

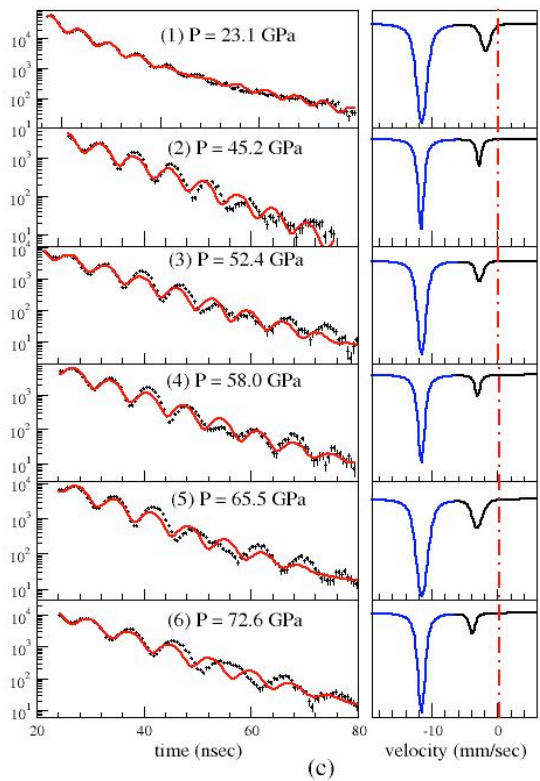

Fig. 7. (Color online) Ambient temperature synchrotron Mössbauer spectra of $^{151}$Eu under pressures in three experimental runs to (a) 25 GPa, (b) 35.6 GPa, and (c) 72.6 GPa in the time domain (left column) and in the corresponding energy domain (right column). Numbers in parentheses give order of measurement. In the left column filled black circles show experimental data with error bars and the red lines show fits to data. In the right column black lines show resonant absorption from Eu sample and the red dash-dotted lines indicate position of zero IS. In (a) blue lines represent the absorption from $Eu_2O_3$ and purple line from EuS as reference. In (b) and (c) blue lines are from EuS as reference.

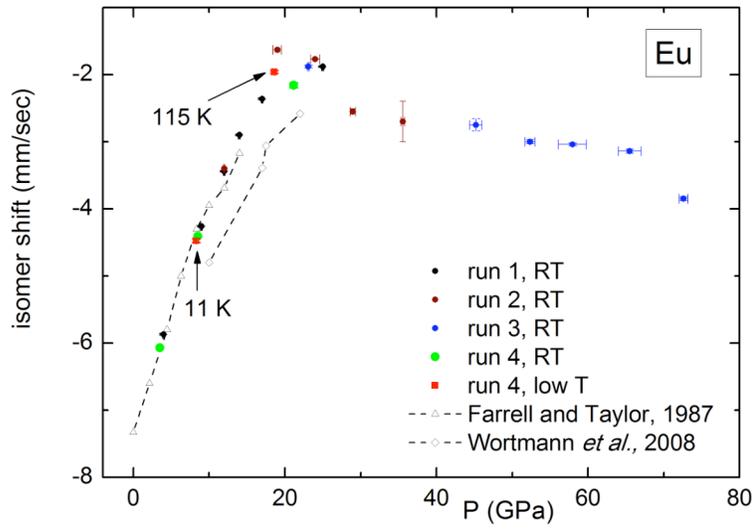

Fig. 8. (Color online) Isomer shift of $^{151}$Eu at ambient temperature and two low temperatures under pressure to 73 GPa in comparison with previous studies [13,15].

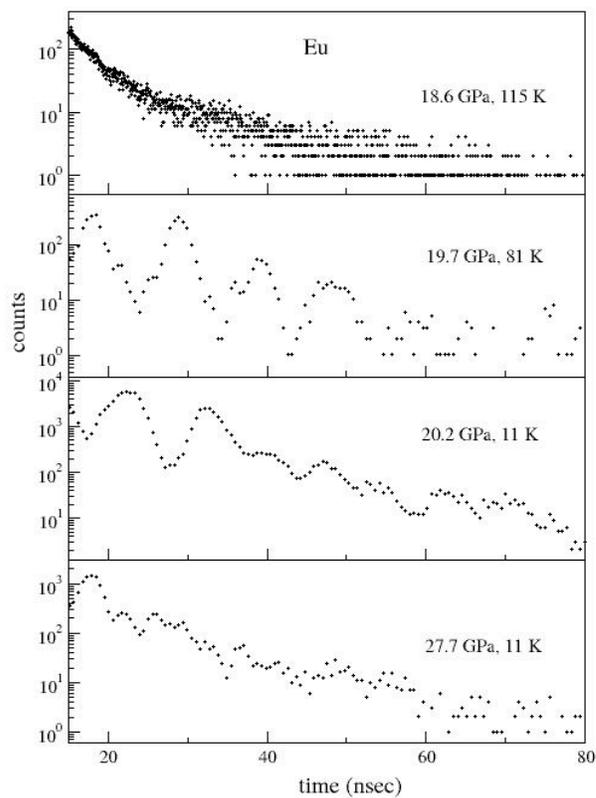

Fig. 9. Synchrotron Mössbauer spectra for Eu metal at low temperatures and high pressures. The oscillations at 27.7 GPa and 11 K, 20.2 GPa and 11 K, and 19.7 GPa and 81 K show Eu orders magnetically, while at 18.6 GPa and 115 K the absence of oscillations shows Eu does not order.

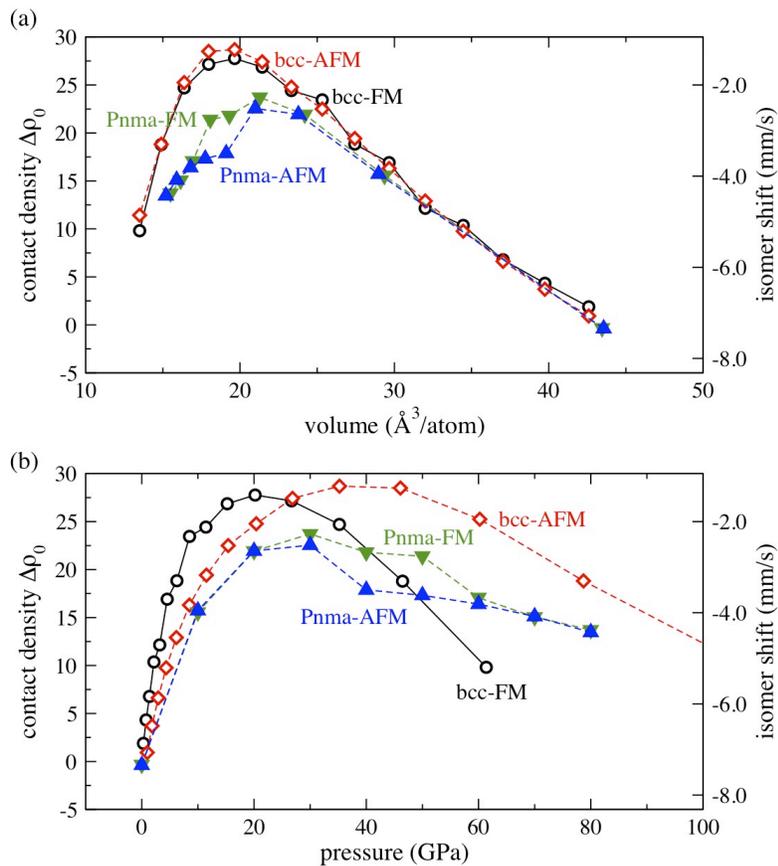

Fig. 10. (Color online) Contact density $\Delta\rho_0$ calculated in DFT as a function of (a) volume and (b) pressure for Eu in the bcc and *Pnma* phase for both ferromagnetic and antiferromagnetic order. The contact density $\Delta\rho_0$ in units electrons/(Bohr radius)$^3$ is shown relative to the contact density of the ferromagnetic bcc phase at zero pressure. The right ordinate shows the linear transformation of the contact density to obtain the isomer shift.

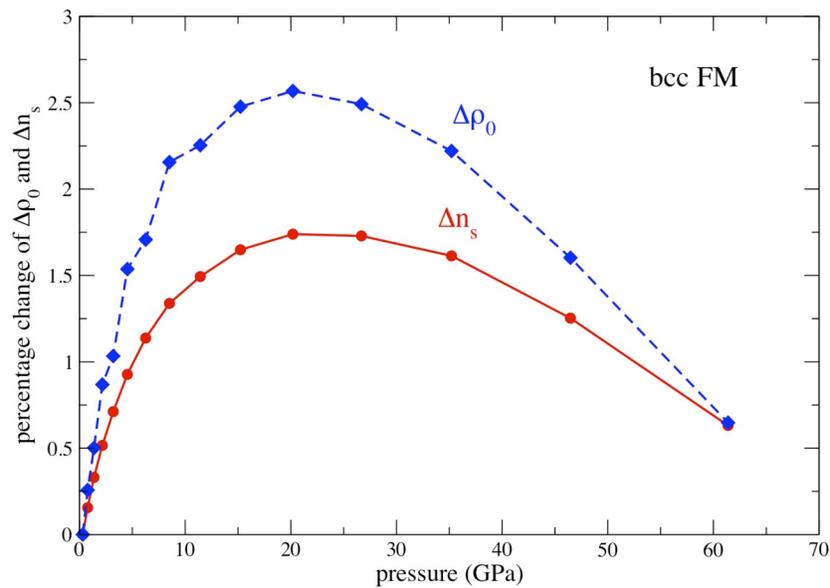

Fig. 11. (Color online) Comparison between the percentage change in the occupation of the valence *s* orbitals of Eu $\Delta n_s$ and the change in contact density $\Delta\rho_0$ (in units electrons/(Bohr radius)$^3$) of the valence electrons for the ferromagnetic bcc Eu phase as a function of pressure. A significant fraction of the change in contact density can be directly attributed to the change in *s*-valence occupation.